\documentstyle[prl,aps]{revtex}

\begin{document}

\draft

\twocolumn[\hsize\textwidth\columnwidth\hsize\csname @twocolumnfalse\endcsname
\title{A Kochen-Specker Theorem for Imprecisely Specified Measurements}
\author{N.\ David Mermin}
\address{Laboratory of Atomic and Solid State Physics, 
Cornell University,  Ithaca, NY 14853-2501}
\maketitle

\begin{abstract}A recent claim that finite precision in the design of real
experiments ``nullifies'' the impact of the Kochen-Specker theorem, is
shown to be unsupportable, because of the continuity of probabilities
of measurement outcomes under slight changes in the experimental
configuration. \end{abstract}

\pacs{PACS numbers: 03.65.Bz, 03.67.Hk, 03.67.Lx}
]

The Kochen-Specker (KS) theorem is one of the major
no-hidden-variables theorems of quantum mechanics. It exhibits a
finite set of finite-valued observables with the following property:
there is no way to associate with each observable in the set a
particular one of its eigenvalues so that the eigenvalues associated
with every subset of mutually commuting observables obey certain
algebraic identities obeyed by the observables
themselves\cite{ft:RMP}.  Such a set of observables is traditionally
called uncolorable.

The physical significance of an uncolorable set of observables stems
from the fact that a simultaneous measurement of a mutually commuting
set must yield a set of simultaneous eigenvalues, which
are constrained to obey the algebraic identities obeyed by the
observables themselves.  So any attempt to assign every observable
in a KS uncolorable set a preexisting {\it noncontextual\/} value (a
``hidden variable'') that is simply revealed by a measurement, will
necessarily assign to at least one mutually commuting subset of
observables a set of values specifying results that quantum mechanics
forbids.  

The term ``noncontextual'' emphasizes that the disagreement with
quantum mechanics only arises if the value associated with each
observable is required to be independent of the choice of the other
mutually commuting observables with which it is measured.  {\it
Contextual\/} value assignments in full agreement with all quantum
mechanical constraints can, in fact, be made.  The import of the KS
theorem is that there exist sets of observables --- ``uncolorable
sets'' --- for which any assignment of preexisting values must be
contextual if all the outcomes specified by those values are allowed
by the laws of quantum mechanics.  The theorem prohibits
noncontextual hidden-variable theories that agree with all the
quantitative predictions of quantum mechanics.

Meyer\cite{ft:Meyer} and Kent\cite{ft:Kent} have questioned the
relevance of the KS theorem to the outcomes of real imperfect
laboratory experiments, by constructing some clever noncontextual
assignments of eigenvalues to every observable in a dense subset of
observables, whose closure contains the KS uncolorable observables.
While no KS uncolorable set can be contained in such a dense colorable
subset, observables in the dense colorable subset can be found
arbitrarily close to every observable in any KS uncolorable set.  This
leads Meyer and Kent to assert that their noncontextual value
assignments to dense sets of observables ``nullify'' the KS theorem.
In support of this claim they note that observables measured in an
actual experiment cannot be specified with perfect precision so, in
Kent's words, ``no Kochen-Specker-like contradiction can rule out
hidden variable theories indistinguishable from quantum theory by
finite precision measurements$\ldots\,$.''\cite{ft:finite}

I show below that this plausible-sounding but not entirely sharply
formulated intuition dissolves under close scrutiny\cite{ft:havlicek}.
First I describe how the KS conclusion that quantum mechanics requires
any assignment of pre-existing values to be contextual can be deduced
directly from the data, even when one is not sure precisely which
observables are actually being measured.  Then I identify where the
intuition of Meyer and Kent goes astray.

At first glance it is not evident that either a KS uncolorable set or
a Meyer-Kent (MK) dense colorable set of observables is relevant when
one cannot specify to more than a certain high precision what
observables are actually being measured.  As traditionally viewed, the
KS theorem merely makes a point about the formal structure of quantum
mechanics, telling us that there is no consistent way to interpret
{\it the theory\/} in terms of the statistical behavior of an
ensemble, in each individual member of which every observable in the
theory has a unique noncontextual value waiting to be revealed by any
appropriate measurement.  Upon further reflection, however, there
emerges a straightforward way to apply the result of the KS theorem to
measurements specified with high but imperfect precision, which makes
it evident that the theorem and its various descendants remain
entirely relevant to imperfect experiments, while the ingenious
constructions of Meyer and Kent do not.

Let us first rephrase the implications of the KS theorem in the ideal
case of perfectly specified measurements.  The theorem gives a finite
uncolorable set of observables, each with a finite number of
eigenvalues.  Because the number of possible assignments of
noncontextual values to observables in the set is finite, no matter
what probabilities are used to associate such values with the
observables, the assignment must give nonzero probability to at least
one mutually commuting subset of the observables having values that
disagree with the laws of quantum mechanics.  So if noncontextual
preexisting values existed and if one could carry out a series of
ideal experiments\cite{ft:state} in each of which one measured a {\it
randomly selected\/} subset of perfectly defined mutually commuting
observables from a KS uncolorable set, then a definite nonzero
fraction of those measurements would produce results violating the
laws of quantum mechanics.  

Conversely, if an appropriately large number of such randomly selected
measurements all yielded results satisfying the relevant quantum
mechanical constraints, then in the absence of bizarre conspiratorial
correlations between one's random choice of which mutually commuting
subset to measure and the hypothetical preexisting noncontextual
values waiting to be revealed by that measurement, one would have
established directly from the ideal data that there could be no
preexisting noncontextual values.

In a real experiment, of course, the observables cannot be precisely
specified.  The actual apparatus used to measure any mutually
commuting subset of an uncolorable finite set of observables will be
slightly misaligned at all stages of the measuring process.  Therefore
the pointer readings from which one deduces their discrete
simultaneous values will give slightly unreliable information about
the ideal observables one was trying to measure.  If the misalignment
is at the limit of ones ability to control or discern, as it will be
in a well designed experiment, then one can and will label the
outcomes of such a procedure with the same discrete eigenvalues used
to label the gedanken outcomes of the ideal perfectly aligned
measurement.  The misalignment will only reveal itself through the
occasional occurrence of runs with outcomes that the laws of quantum
mechanics prohibit for a perfectly aligned
apparatus\cite{ft:original}. But although the outcomes deduced from
such imperfect measurements will occasionally differ dramatically from
those allowed in the ideal case, if the misalignment is very slight,
{\it the statistical distribution of outcomes will differ only
slightly from the ideal case.}

It is this continuity of quantum mechanical probabilities under small
variations in the experimental configuration (without which quantum
mechanics, or, for that matter, any other physical theory, would be
quite useless) that makes the KS conclusion relevant to the imperfect
case.  Even though the apparatus cannot be perfectly aligned, if
quantum mechanics is correct in its quantitative predictions, then the
fraction of runs which violate the quantum-mechanical rules applying
to the ideal observables can be made {\it arbitrarily small\/} in the
realistic case by making the alignment sufficiently sharp.  But the KS
theorem tells us that if the possible results of the ideal gedanken
measurements were consistent with preexisting {\it noncontextual
values}, then the fraction of quantum-mechanically forbidden outcomes
for the real experiments would have to approach a {\it nonzero
limit\/} as the alignment became sharp.

By making the experimental misalignment sufficiently small, one can
make the statistics of the slightly unreliable results of the randomly
selected realistic measurements arbitrarily close to the statistics of
the theoretical results of the randomly selected ideal measurements.
Therefore a failure in the realistic case to observe {\it sufficiently
many\/} values that contradict the constraints imposed on the data by
quantum mechanics can demonstrate, just as effectively as the failure
to observe {\it any\/} such contradictions in the ideal case, that if
the measurements are revealing preexisting values, then those values
must be {\it contextual\/}.  

Because it can be stated in terms of outcome probabilities, and
because those probabilities must vary continuously with variations in
the experimental apparatus, the conclusion of the KS theorem are not
``nullified'' by the finite precision with which actual measurements
can be specified.

\bigskip

But what about Meyer's and Kent's intuition that the dense colorable
MK set can be used to furnish a group of nearly ideal experiments with
noncontextual values that agree with the constraints imposed by
quantum mechanics on all mutually commuting ideal subsets?  The
crucial word here is ``all''.  It is impossible for a colorable set of
observables in one-to-one correspondence with the observables in a KS
uncolorable set, to have mutually commuting subsets that correspond to
{\it every\/} mutually commuting subset of the KS uncolorable set.  At
least one of those subsets cannot be mutually commuting\cite{ft:case}.

This impossibility is established by the KS theorem itself, which uses
only the topology of the network of links between commuting
observables in the full KS uncolorable set.  This topology would be
preserved by the correspondence between the nearby colorable and
uncolorable sets, if the correspondence between their mutually
commuting subsets were complete.  Because, as Meyer and Kent
explicitly show, the MK set is colorable, the correspondence cannot be
complete.  

Thus any finite MK colorable set in sufficiently close one-to-one
correspondence with a finite KS uncolorable set, must necessarily lack
the full range of mutually commuting subsets that the KS uncolorable
one contains. There is therefore at least one mutually commuting
subset of the KS uncolorable set for which the MK colorable set fails
to provide a set of values agreeing with the constraints imposed by
quantum mechanics.  It is only this deficiency that makes it possible
to color the observables in the MK set in a noncontextual way that
satisfies the quantum mechanical constraints for all mutually
commuting subsets.  But this same deficiency makes the MK set useless
for specifying preassigned noncontextual values agreeing with quantum
mechanics for the outcomes of every one of the slightly imperfect
experiments that corresponds to measuring a mutually commuting subset
of observables from the ideal KS uncolorable set.

If one tries to bridge this gap in the argument by associating more
than a single nearby MK colorable observable with some of the
observables in the ideal uncolorable set, one sacrifices the
noncontextuality of the value assignments.  Nor does the MK
``nullification'' of the KS theorem work in a toy universe in which
the physically allowed observables are restricted to be those in the
dense colorable set.  In such a universe measurements still could not
be specified with perfect precision, and one would still have to rely
on the continuity of outcome probabilities with small changes in the
experimental configuration to relate the theory to actual imperfect
observations.  Under such conditions it would be highly convenient to
introduce fictitious observables which were the limit points of
physical observables in the dense colorable set, whose statistics
could represent to high accuracy the statistics of physically allowed
observables in their immediate vicinity.  The KS theorem would hold
for these fictitious limit-point observables, and would therefore
apply by continuity to measurements of the nearby physical
observables, for exactly the reasons I have just described in the case
of conventional quantum mechanics with its continuum of observables.
Indeed, I can see no grounds other than convenience (which is, of
course, so enormous as to be utterly compelling) for treating the
whole continuum of observables as physically real, as we actually do,
rather than regarding it as a fictitious extension of some countable
dense subset of observables.

So contrary to the claim of Meyer and Kent, the KS theorem is not
nullified by the finite precision of real experimental setups because
of the fundamental physical requirement that probabilities of outcomes
of real experiments vary only slightly under slight variations in the
configuration of the experimental apparatus, and because the import of
the theorem can be stated in terms of whether certain outcomes never
occur, or occur a definite nonzero fraction of the time in a set of
randomly chosen ideal experiments.  

But the elegant MK colorings of dense sets of observables make an
instructive contribution to our understanding of the KS theorem, by
forcing us to recognize that a principle of continuity of physical
probability, obeyed by the quantum theory, plays an essential role in
relating the conclusions of the theorem to real experiments.  Indeed,
the relationship between the MK dense colorable sets and the KS
uncolorable set offers a novel perspective on why it is sensible to
base physics on real (as opposed to rational) numbers, in spite of the
finite precision of actual experimental arrangements.

\bigskip

{\bf Acknowledgment.}  This reexamination of the physical setting of
the Kochen-Specker theorem was supported by the National Science
Foundation, Grant No. PHY 9722065.


\begin{references}
\bibitem{ft:finite}
``Finite precision'' refers not to the
reliability of the measurement outcome, but to what it is that is
actually being measured.
\bibitem{ft:state} Because the KS theorem does not depend on the state
in which the system is prepared, the ensemble of systems on which the
experiments are performed need not contain identically prepared
systems, nor need the states of the systems be pure states.
\bibitem{ft:original} 
In the original KS argument, for example, one
measures the mutually commuting squares of three orthogonal spin
components of a spin-1 particle, and the laws of quantum mechanics
guarantee that one must record two 1's and one 0.  If, however, the
apparatus is imperfect and the three directions are not precisely
orthogonal, one will occasionally get different numbers of 1's and
0's, in apparent violation of the laws for spin-1.  (If the three
directions are not precisely orthogonal, the relevant three
observables will not exactly commute, so the statistics of the
``measurement'' will in general depend slightly on how it is carried
out.)
\bibitem{ft:case} In the case of squared spin-1 components considered
by Meyer, for example, it is impossible to satisfy the condition of
mutual orthogonality for every trio of components in the MK colorable
set approximating a mutually orthogonal trio of components in the KS
uncolorable set.  But a set of squared spin-components in the MK set
that is not associated with exactly three orthogonal directions, need
not satisfy the quantm-mechanical requirement that their assigned
values consist of two 1's and one 0.
\end{references}
\end{document}